\documentclass{article}
\usepackage{spconf,amsmath,graphicx,hyperref}
\usepackage{graphicx,verbatim}
\usepackage{amsmath}
\usepackage{enumerate}
 \usepackage{hyperref}
\usepackage{mathrsfs}
\usepackage{bm}
\usepackage{xspace}
\usepackage{graphicx}
\usepackage{float} 
\usepackage{url}
\usepackage{subcaption}
\usepackage{tabularray}

\title{Kalman Filter Based Linear Deformable \\ for Retinal Vessel Segmentation}
\name{Zhihao Zhao$^{a}$\thanks{This research is supported by Bavarian State(AZ-1503-21).
Email: zhihao.zhao@tum.de},Yinzheng Zhao$^{a}$,Junjie Yang$^{a}$ 
,Quanmin Liang$^{b}$,Daniel Zapp$^{a}$,Kai Huang$^{b}$,M.Ali Nasseri$^{a,c,d}$ }

\address{$^{a}$ TUM Klinikum Rechts der Isar, Technische Universität München, München, Germany  \\
      $^{b}$ School of Computer Science and Engineering, Sun Yat-Sen University, Guangzhou, China \\ 
      $^{c}$ Zhongshan Ophthalmic Center, Sun Yat-sen University, Guangzhou, China\\
      $^{d}$ University of Alberta, Alberta, Canada \\}

\begin{document}
%
\maketitle
\newcommand{\KaLDeX}{{\textbf{KaLDeX}}\xspace}
\begin{abstract}

In the realm of ophthalmic imaging, accurate vascular segmentation is paramount for diagnosing and managing various eye diseases. 
Contemporary deep learning-based vascular segmentation models rival human accuracy but still face substantial challenges in accurately segmenting minuscule blood vessels in neural network applications.
In this paper, we propose a novel network (\KaLDeX) for vascular segmentation leveraging a Kalman filter based linear deformable cross attention (LDCA) module integrated within a UNet$++$ framework. 
Our approach is  based on two key components: Kalman filter (KF) based linear deformable convolution (LD) and cross-attention (CA) modules. 
The LD module is designed to adaptively adjust the focus on thin vessels that might be overlooked in standard convolution. 
The CA module improves the global understanding of vascular structures.
Finally, we adopt a topological loss function based on persistent homology to constrain the topological continuity of the segmentation.
The proposed method is evaluated on retinal fundus image datasets (DRIVE, CHASE\_BD1, and STARE) as well as the 3 mm and 6 mm of the OCTA-500 dataset, achieving an average accuracy (ACC) of 97.25\%, 97.77\%, 97.85\%, 98.89\%, and 98.21\%, respectively.

\end{abstract}

\begin{keywords}
Vessel Segmentation, Linear Deformable, Cross-Attention, Kalman Filter
\end{keywords}

\section{Introduction}
\label{sec:intro}

The accurate segmentation of the retinal vasculature is essential in diagnosing and treating a myriad of eye diseases \cite{xu2020retinal, guo2020dense, yu2022vision}. 
The segmentation of retinal vessels is challenging due to their small, complex, and highly tortuous structures with varied shapes. In standard UNet \cite{zhang2021pyramid, guo2021channel} architectures, the downsampling within convolutional modules leads to the loss of this fine-grained information. This is particularly problematic for retinal vessel segmentation, especially resembling the small vascular structures. Furthermore, traditional feature extraction methods using standard convolutions struggle to track the irregular paths of vessels, often resulting in discontinuous segmentations.
To address the information loss from downsampling in a standard UNet, UNet++ \cite{zhou2018unet++} employs a strategy involving multiple nested skip connections to help capture feature information from different layers. However, while the nested skip connections in UNet++ help preserve original image features, they still cannot guarantee the accurate identification of vessel regions that are highly complex and varied in shape. On the other hand, conventional deformable convolution, which dynamically adjusts its receptive field, is effective at capturing the spatial contours of large, irregularly shaped objects. Yet, the linear structure of blood vessels represents an extreme form of irregularity, which remains difficult for even standard deformable convolutions to handle effectively.

DSCNet \cite{qi2023dynamic} introduced a method known as dynamic snake convolution to resolve the challenge that standard deformable convolution faces with linear structures. This approach reshapes the convolutional kernel into a one-dimensional, snake-like form for the convolution operation. Concurrently, the network predicts an offset field corresponding to this 1D kernel. The method operates sequentially. It initializes the offset at the kernel's center to zero, and the offset for each subsequent point is predicted relative to the position of the preceding one. This strategy has proven effective for processing linear structures, as a larger kernel can capture vessels more comprehensively and reduce discontinuities.
However, this design has a significant drawback. The accumulation of prediction errors. Because the prediction for each point is conditioned on the previous one, errors are propagated and amplified along the kernel. Consequently, points located farther from the kernel's center are subject to a larger cumulative localization error.

To address these limitations, we introduce a novel approach KaLDeX, that enhances retinal vessel segmentation accuracy by leveraging a Kalman filter (KF) \cite{welch1995introduction}-based linear deformable cross-attention mechanism. Our network includes two innovative modules: a KF-based linear deformable convolution module (LD) and a cross-attention module (CA). The KF-based LD module employs an automatic update method that iteratively predicts the position of the next optimal point, starting from the center of the kernel. Conceptually, the KF can be understood as a dynamic weighted averaging process that incorporates historical information. It leverages the inherent prior that vascular structures are continuous and locally smooth. By fusing coordinate information and giving greater weight to the more reliable predictions near the kernel's center, this approach mitigates the cumulative error in the network's predicted offsets, particularly at the periphery of the kernel. Furthermore, to effectively integrate the global vascular information captured by UNet++ with the local geometric variations perceived by our LD module, we employ a cross-attention mechanism to fuse these two distinct sets of features.  By aggregating the detailed features captured by the LD module with the global information obtained through the feature map in UNet++, the network gains a more comprehensive understanding of the vascular network. To ensure the continuity and topological structure of vascular structures, we employ the Centerline Dice (clDice) loss function and the optimized topological loss function to reveal the topological continuity of the segmentation structure. This approach not only enhances the fidelity of vessel segmentation but also preserves the geometric integrity of vascular structure.

Our contributions can be summarized as follows. Firstly, we propose a Kalman filter based linear deformable module suitable for vascular segmentation in ophthalmic images, which enables the convolutional field of view to focus more on the vascular regions. Secondly, we aggregate feature maps captured by LD modules with the  contextual information obtained through UNet++ to enhance the network's ability to discern small vascular structures. Finally, we integrate Kalman filter based linear deformable cross-attention into the UNet++ framework. Empirical evidence demonstrates that our method surpasses the capabilities of the current best models in retina  vessel segmentation.


\section{PROPOSED METHOD}

\newcommand{\R}{{\bm{$R$}}\xspace}
\label{sec:format}

\begin{figure*}[ht]
	\centering
	\includegraphics[width=.85\textwidth]{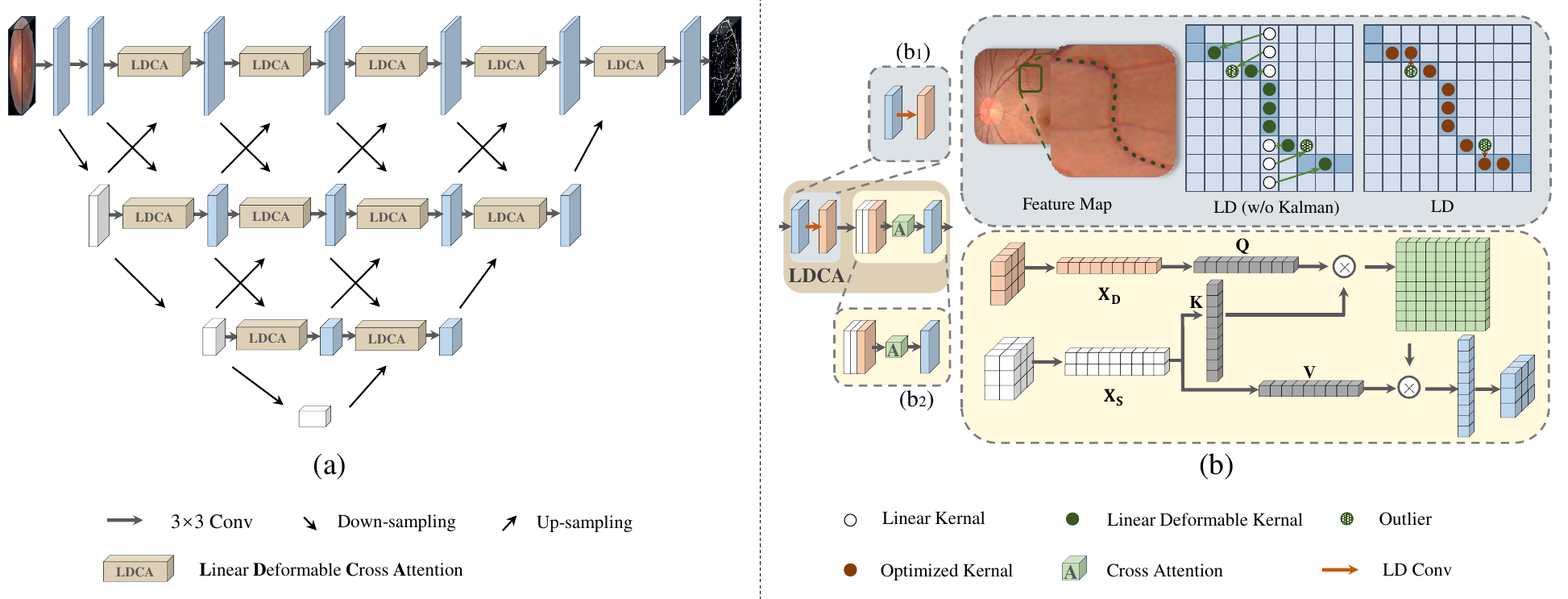}
	\caption{Overview of our proposed \KaLDeX. (a) is the overall framework of our proposed structure. (b) showcases the key module (LDCA), which primarily consists of two sub-modules: LD($b_1$) and CA($b_2$). LD is primarily focused on capturing vascular structures through deformable convolutions and optimizing the coordinates of deformable convolutions using Kalman filtering. CA aggregates the output feature maps of LD with the feature maps obtained from up-sampling and down-sampling in the network.} 
	\label{fig:flowchart}
\end{figure*}

\subsection{Structure Overview}
Fig.~\ref{fig:flowchart}(a) represents the overall structure of our \KaLDeX model. The network utilizes UNet++ as our foundational backbone, comprising convolutional layers, up-sampling and down-sampling layers, along with our Kalman filter based linear deformable cross attention (LDCA) module.
Fig.~\ref{fig:flowchart}(b) showcases our proposed LDCA module, which primarily consists of two sub-modules: Kalman filter based Linear Deformable (LD) Convolution  and Cross-Attention (CA). Fig.~\ref{fig:flowchart}($b_1$) is the LD sub-module,  which is primarily focused on capturing vascular structure information more effectively through deformable convolution. Simultaneously, we employ Kalman filtering to reduce the accumulation of errors in convolutional kernel coordinate offsets during the execution of deformable convolutions. This optimization aims to obtain a feature map from linear deformable convolutions that closely represents the vascular structure in the acquired field of view.
Fig.~\ref{fig:flowchart}($b_2$) illustrates that the CA sub-module enhances the model’s ability to discern small vessels and understand the vascular structure context by aggregating the output feature map of LD with the feature maps obtained from up-sampling and down-sampling in UNet++.

\subsection{Kalman Filter based Linear Deformable Convolution}

Our linear deformable convolution (LD) structure is inspired by deformable convolution networks (DCNs) and dynamic snake convolution networks (DSCNet) \cite{qi2023dynamic}.  The linear convolutions employed are one-dimensional linear convolution kernels sized $9\times1$ and $1\times9$, respectively. Our discussion will focus solely on the horizontal coordinates, as the vertical coordinates are identical. Each convolution kernel is represented as $Ker=(x_{i\pm c})$, where $c={0,1,2,3,4}$. DSCNet uses a learnable offset $\delta$ to predict the deviation in the coordinates of the deformed convolution kernel $x_{i}=x_{i-1}+\delta_{i}$. Each new coordinate is the sum of the previous coordinate and the predicted deviation, thus resulting in a cumulative errors.
The Kalman filter is a highly effective method for reducing cumulative errors when predicting future values by weighting the current value and previous values \cite{welch1995introduction}.
The principal concept of our LD module involves assigning a weight $K$ to each offset $\delta_{i}$, with the new update formula $x_i = (1-K_i)x_{i-1}+K_{i}(x_{i-1}+\delta_{i})$. This step originates from the Kalman filter update formula presented in Equation~\ref{equ:kal_update}. Here, $K_{i}$ represents the Kalman filter gain, which can be iteratively derived from Equation~\ref{equ:kal_update}.
In the equation, the parameters $p$ represent the estimated covariance. $r$ is a hyperparameter, which is associated with the errors present in the measurements from the neural network output.

The hyperparameter $r$ is empirically set to $0.1^2$. The initial values of $p_0$ and $x_0$ are set to $1$ and $0$ respectively. These parameters will be updated iteratively during the process. 
\begin{equation}
\begin{cases}
    K_{\mathrm{i}}=\frac{p_{i-1}}{p_{i-1}+r}\\
    x_i=(1-K_i)x_{i-1}+K_i(x_{i-1}+\delta_{i})=x_{i-1}+K_i\delta_{i}\\
    p_{i}=(1-K_i)p_{i-1}
\end{cases}   
\label{equ:kal_update}
\end{equation}

\subsection{Cross Attention Aggregation Module}

Inspired by the advantage that cross-attention mechanisms can capture the associative information between different positions within two input sequences
The CA module in our work is utilized to aggregate the vascular detail information obtained from the LD module and the vascular contextual features derived from up-sampling and down-sampling in UNet++. As illustrated in Fig.~\ref{fig:flowchart}($b_2$) In the color figure, the feature maps obtained by regular convolution are represented in blue, the feature maps resulting from the LD module are represented in orange, while the feature maps in white represent the results from the up-sampling and down-sampling in UNet++. 
Initially, we need to flatten the feature maps into one dimension. Then, the result $X_D$ of the LD module is transformed into $Q$ matrix through a linear connection layer $W_Q$, while the up-sampling and down-sampling results $X_S$ of UNet++ are transformed into $K$ and $V$ matrices through $W_K$ and $W_V$, respectively. The $Q$ and $K$ matrices are multiplied to create an attention map. This attention map is then multiplied with the $V$ matrix to yield the aggregated result. Finally, the output is reshaped to match the shape of the input image. 
Overall, the entire aggregation process can be defined by Equation~\ref{eq:ca}, where $d_k$ is the dimension of $K$.
\begin{equation}
\begin{aligned}
   \mathrm{CA}(X_D,X_S)&=\mathrm{softmax}\left(\frac{Q{K}^{\top}}
    {\sqrt{d_k}}\right)V\\
    Q = W_QX_D, K &= W_KX_S, V = W_VX_S\\
\end{aligned}    
\label{eq:ca}
\end{equation}

\subsection{Loss Function}
In the segmentation of vascular structures, besides focusing on small vascular details, the overall continuity in the vascular space is also crucial. To maintain the topological continuity in the segmentation results, we introduce a clDice loss based on topological structure similarity as proposed in \cite{shit2021cldice}. We incorporate this topological similarity along with the cross-entropy loss function to form our new loss. The topological similarity measurement is defined according to Equation~\ref{eq:dice}, where $V_L$ is the ground truth of vascular segmentation, $V_P$ is the predicted result, $S_L$ is the skeleton extracted from the ground truth, and $S_P$ is the skeleton extracted from the predicted result. The topological accuracy $\mathrm{Tprec}(S_P, V_L)$ and topological sensitivity $\mathrm{Tsens}(S_L, V_P)$ can be defined accordingly.
\begin{equation}
\begin{aligned}
   \mathrm{Tprec}(S_P,V_L)=\frac{|S_P\cap V_L|}{|S_P|},\mathrm{Tsens}(S_L,V_P)=\frac{|S_L\cap V_P|}{|S_L|}\\
    \mathrm{clDice}(V_P,V_L)=2\times \frac{\mathrm{Tprec}(S_P,V_L)\times \mathrm{Tsens}(S_L,V_P)}{\mathrm{Tprec}(S_P,V_L)+\mathrm{Tsens}(S_L,V_P)}\\
\end{aligned}    
\label{eq:dice}
\end{equation}

Finally, the weighted sum of the clDice and cross-entropy loss functions is employed as the loss function:
\begin{equation}
    \mathcal{L}_{oss}=\alpha(1-clDice)+ (1-\alpha) \mathcal{L}_{BCE}
    \label{eq:loss}
\end{equation}


\section{EXPERIMENTS}
\label{sec:pagestyle}

\newsavebox\CBox
\def\textBF#1{\sbox\CBox{#1}\resizebox{\wd\CBox}{\ht\CBox}{\textbf{#1}}}
\subsection{Setup Details}

\subsubsection{Datasets.}
All experiments are carried out on publicly available Ophthalmic datasets. 
The fundus image datasets include STARE \cite{hoover2000locating}, DRIVE \cite{staal2004ridge} and CHASE\_DB1 \cite{chasedb1}. The OCTA-500 \cite{li2024octa} which comprises 500 OCTA projection images: 300 images with a 6 mm × 6 mm field of view and 200 images with a 3 mm × 3 mm field of view, all provided with corresponding annotations.


\subsubsection{Experimental Settings.}
Our experimental setup uses a single NVIDIA RTX A5000 GPU (24 GB) with PyTorch. The learning rate is initially 0.0025. The dataset is trained for 10 epochs. To reduce overfitting, online augmentation techniques like flipping and Gaussian noise (5x5 kernel) are applied.  

\begin{figure}[ht]
    \centering
    \includegraphics[width=0.48\textwidth]{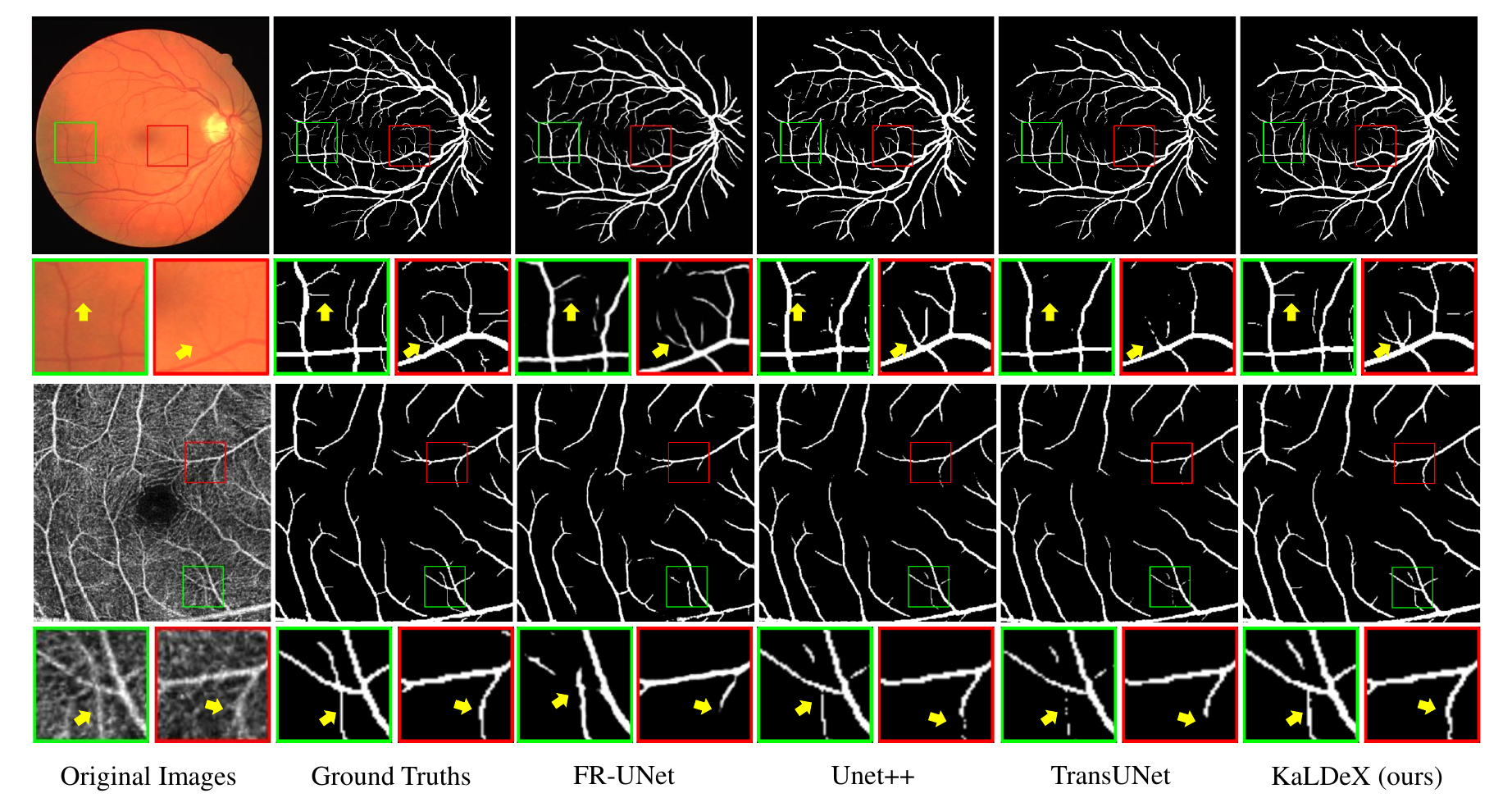}
    \caption{Visualization of segmentation results. The first and third rows depict the segmentation results of fundus images and OCTA, respectively, while the second and fourth rows display magnifications of the cropped regions.}
    \label{fig:segment}
    
\end{figure}

\subsection{Evaluation and Results}
\subsubsection{Results of Segmentation}
In this study, we comprehensively evaluated our vascular segmentation experiments and compared our method with leading approaches, including UNet \cite{ronneberger2015u}, UNet++ \cite{zhou2018unet++}, FR-UNet \cite{liu2022full}, TransUNet \cite{chen2021transunet}, and RVSeg \cite{wang2020rvseg}. Fig.~\ref{fig:segment} presents the segmentation results of our model alongside traditional architectures, highlighting their advantages and limitations. A close examination of the cropped regions reveals that many small and faint vascular structures, which are crucial for accurate diagnosis, are often missed or inadequately captured by traditional networks. 
In contrast, our network excels at identifying and segmenting small vessels, capturing intricate vascular details that traditional methods frequently overlook.

\begin{table}[th]
\centering
\caption{Comparison with the State-of-The-Art Methods On DRIVE AND CHASE\_DB1 }
\label{tab:drivechase}
\resizebox{0.45\textwidth}{!}{%
\begin{tblr}{
  cell{1}{1} = {r=2}{},
  cell{1}{2} = {c=6}{c}, %
  cell{1}{8} = {c=6}{c},
  vline{2,8} = {1-14}{},
  hline{1,3,8,9} = {-}{},
  hline{2} = {2-14}{},
}
Methods        & DRIVE (\%)  &        &        &        &        &        & CHASE\_DB1 (\%) &     &         &        &       &        \\
    & Acc    & Sen    & Spe    & AUC    & Dice     & IOU    & Acc    & Sen    & Spe    & AUC    & Dice     & IOU    \\
U-Net \cite{ronneberger2015u}     & 96.78 & 80.57 & 98.33 & 98.25 & 81.41 & 68.64 & 97.43 & 76.50 & \textBF{98.84} & 98.36 & 78.98 & 65.26 \\
UNet++ \cite{zhou2018unet++}    & 96.79 & 78.91 & 98.50 & 98.25 & 81.14 & 68.27 & 97.39 & 83.57 & 98.32 & 98.81 & 80.15 & 66.88 \\
FR-UNet \cite{liu2022full}   & 97.05 & \textBF{83.56} & 98.37 & 98.89 & \textBF{83.16} & \textBF{71.20} & 97.48 & \textBF{87.98} & 98.14 & \textBF{99.13} & 81.51 & 68.82 \\
TransUNet \cite{chen2021transunet}  & 96.62 & 79.06 & 98.31 & 97.74 & 80.39 & 67.21 & 97.30 & 83.84 & 98.20 & 98.48 & 79.64 & 66.17 \\
RVSeg  \cite{wang2020rvseg} & 96.81 & 81.07 & 98.45 & 98.17 & 81.07      & 68.17      & 97.26 & 80.69 & 98.36 & 98.33 & 79.02      & 65.32      \\
LDCANet(ours)   & \textBF{97.25} & 81.34 & \textBF{98.69} & \textBF{98.92} & 82.87 & 70.78 & \textBF{97.77} & 83.49 & 98.69 & 99.11 & \textBF{81.72} & \textBF{69.12}           
\end{tblr}
}
\end{table}

Our evaluation employs metrics such as AUC, Acc, and the DICE coefficient to quantitatively assess our model’s superior performance. Table~\ref{tab:drivechase} provides a detailed comparison of the DRIVE and CHASE\_DB1 datasets, demonstrating our model’s effectiveness in vascular segmentation. Similarly, Table~\ref{tab:stare} showcases comparative performance metrics on the STARE dataset, reinforcing our model's adeptness in this domain, while Table~\ref{tab:octa} delineates our model's segmentation prowess on the OCTA dataset, further substantiating the versatility and robustness of our proposed structure across varied imaging modalities.
Tables~\ref{tab:drivechase} and ~\ref{tab:stare} confirm that our model sets a new benchmark in accuracy (ACC), achieving 97.25\%, 97.77\%, and 97.85\% on the DRIVE, CHASE\_DB1, and STARE datasets, respectively. Beyond accuracy, our method excels across multiple evaluation metrics, as detailed in the concluding rows of Tables~\ref{tab:drivechase} and ~\ref{tab:stare}.

For the OCTA dataset, we focused on ACC and the DICE coefficient. Table~\ref{tab:octa} highlights our model’s top performance on OCTA\_3mm and OCTA\_6mm datasets, demonstrating its adaptability and effectiveness in addressing the complexities inherent in segmenting vascular structures across diverse datasets.

\begin{table}[t]
\begin{minipage}{0.238\textwidth}
    \caption{Quantification \\on STARE}
    \label{tab:stare}
    \resizebox{\textwidth}{!}{%
    \begin{tblr}{
      cell{1}{1} = {r=2}{},
      cell{1}{2} = {c=6}{c}, %
      vline{2} = {1-10}{},
      hline{1,3,8,9} = {-}{},
      hline{2} = {2-7}{},
    }
    Methods          & STARE (\%)  &        &        &        &        &        \\
                     & Acc    & Sen    & Spe    & AUC    & Dice     & IOU \\
    U-Net \cite{ronneberger2015u}       & 97.30 & 78.50 & 98.84 & 98.75 & 81.18 & 68.56 \\
    UNet++ \cite{zhou2018unet++}      & 96.34 & 70.09 & 98.83 & 98.84 & 81.50 & 69.02 \\
    FR-UNet \cite{liu2022full}     & 97.52 & \textBF{83.27} & 98.69 & \textBF{99.14} & \textBF{83.30} & \textBF{71.56} \\
    TransUNet \cite{chen2021transunet}   & 97.27 & 77.87 & 98.98 & 98.99 & 81.06 & 69.97 \\
    RVSeg \cite{wang2020rvseg}      & 97.19 & 77.57 & 98.93 & 99.03 & 82.78      & 69.38      \\
    LDCANet(ours)    & \textBF{97.85} & 78.61 & \textBF{99.18} & 99.01 & 82.20 & 70.06            
    \end{tblr}
    }
\end{minipage}
\begin{minipage}{0.236\textwidth}
    \caption{ Quantification \\on OCTA}
    \label{tab:octa}
    \resizebox{\textwidth}{!}{%
    \begin{tblr}{
      cell{1}{1} = {r=2}{},
      cell{1}{2} = {c=2}{c}, %
      cell{1}{4} = {c=2}{c},
      vline{2,4} = {-}{},
      hline{1,3,7,8} = {-}{},
      hline{2} = {2-6}{},
    }
    Methods        & OCTA\_3mm (\%)  &     & OCTA\_6mm (\%) &      \\
                   & Acc    & Dice   & Acc   & Dice    \\
    U-Net \cite{ronneberger2015u}      & 95.45 & 88.35 & 95.21 & 85.03 \\
    U-Net++ \cite{zhou2018unet++}    & 95.98 & 88.64 & 95.73 & 85.67 \\
    FR-UNet \cite{liu2022full}     & 98.84 & 91.15 & 98.02 & 88.85 \\
    TransUNet \cite{chen2021transunet}    & 96.32 & 90.89 & 97.42 & 88.67 \\
    LDCANet(ours) & \textBF{98.89} & \textBF{91.59} & 98.21 & 89.13           
    \end{tblr}
    }
\end{minipage}
\end{table}

\subsubsection{Ablation Study.} 
To assess the efficacy of each proposed module, we conducted ablation experiments, with results in Table~\ref{tab:ablation}. These studies examined the impact of the LD and CA modules. Notably, adding the linear deformable module significantly improved segmentation, boosting accuracy by nearly 2\% on the OCTA dataset. Further optimization of its convolutional field of view using Kalman filtering also had a positive effect. Lastly, the Cross-Attention module enhanced feature aggregation, further increasing accuracy across datasets.
\begin{table}[th]
\centering
\caption{Ablation study of the proposed LDCA module}
\label{tab:ablation}
\resizebox{0.5\textwidth}{!}{%
\begin{tblr}{
  cell{1}{1} = {r=2}{},
  cell{1}{2} = {c=2}{c}, %
  cell{1}{4} = {c=2}{c}, %
  cell{1}{6} = {c=2}{c}, %
  cell{1}{8} = {c=2}{c}, %
  cell{1}{10} = {c=2}{c},
  vline{2,4,6,8,10} = {1-6}{},
  hline{1,3,7} = {-}{},
  hline{2} = {2-12}{},
}
Methods                             & DRIVE (\%)  &        & CHASE\_DB1 (\%) &     & STARE (\%)  &         & OCTA\_3mm (\%) &     & OCTA\_6mm (\%) &      \\
                                  & Acc    & Dice   & Acc    & Dice   & Acc    & Dice    & Acc   & Dice   & Acc   & Dice     \\

Baseline                 & 96.79 & 81.40 & 97.41 & 80.42 & 97.34 & 81.59 & 96.18 & 88.67 & 96.32 & 85.81  \\
Baseline+LD(w/o Kalman)  & 97.04 & 82.65 & 97.58 & 81.48 & 97.45 & 81.78 & 98.12 & 91.04 & 98.02 & 88.05  \\
Baseline+LD(w Kalman)    & 97.08 & 82.71 & 97.59 & 81.51 & 97.50 & 81.78 & 98.88 & 91.15 & 98.12 & 88.85  \\
Baseline+LDCA(w Kalman)  & 97.25 & 82.87 & 97.77 & 81.72 & 97.85 & 82.20 & 98.89 & 91.59 & 98.21 & 89.13  \\
\end{tblr}
}
\end{table}


\section{CONCLUSION}
\label{sec:typestyle}

In this paper, we propose KaLDeX, a novel network leveraging a Kalman-based linear deformable cross-attention mechanism to improve  the segmentation accuracy of vascular structures in ophthalmic images. By integrating a Kalman filter-based linear deformable convolution module with a cross-attention module, our approach refines vascular focus and enhances small structure discernment. Embedded in the UNet++ framework, the LDCA module merges detailed features with broad contextual information, enabling a comprehensive understanding of the vascular network. Our method introduces innovative advancements in medical image analysis, significantly improving segmentation accuracy.

\vfill\pagebreak




\bibliographystyle{IEEEbib}
\bibliography{strings,refs}

\end{document}